\documentclass{article}
\usepackage{spconf,amsmath,epsfig}
\usepackage{amsf onts,amssymb}
\usepackage{subfigure}
\usepackage{stfloats}
\usepackage{stmaryrd}
\usepackage{float}
\newcommand{\ba}{\left( \begin{array}}
\newcommand{\ea}{\end{array} \right)}
\newcommand{\bq}{\begin{eqnarray*}}
\newcommand{\eq}{\end{eqnarray*}}
\newcommand{\bqn}{\begin{eqnarray}}
\newcommand{\eqn}{\end{eqnarray}}

\newtheorem{theorem}{Theorem}

\usepackage{hyperref}
\hypersetup{
    colorlinks=true,%
    citecolor=black,%
    filecolor=blue,%
    linkcolor=red,%
    urlcolor=black,}

\usepackage[normalem]{ulem}

\title{Sulcal Pattern Matching with the Wasserstein Distance}
\name{
Zijian Chen, Soumya Das, Moo K. Chung\thanks{Published in Proceedings of the 2023 IEEE International Symposium on Biomedical Imaging (ISBI). The study is funded by NIH EB022856, EB02875, NSF MDS-2010778.}
}
\address{University of Wisconsin, Madison, USA\\
{\tt \{zijian.chen,mkchung\}@wisc.edu}}

\begin{document}

\maketitle

\begin{abstract} 
We present the unified computational framework for modeling the sulcal patterns of human brain obtained from the magnetic resonance images. 
The Wasserstein distance is used to align the sulcal patterns nonlinearly. These patterns are topologically different across subjects making the pattern {matching a challenge.} 
We work out the mathematical details {and develop the gradient  descent algorithms for estimating the deformation field. We {further} quantify the image registration performance. This method is applied in identifying the differences between male and female sulcal patterns.} 
\end{abstract}

\section{Introduction}

\begin{figure}[t]
	\centering
	\includegraphics[width=\linewidth,clip=true]{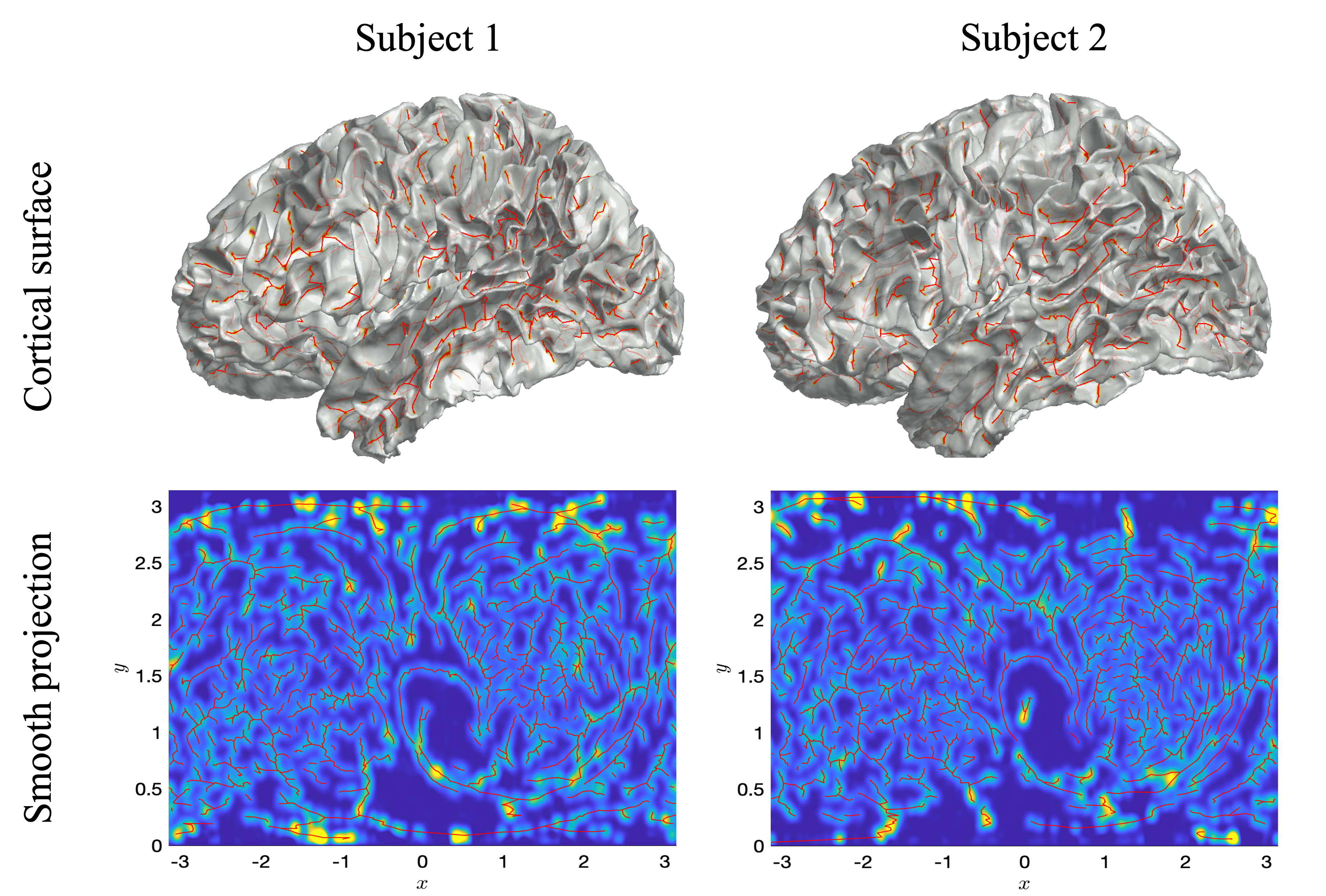}
	\caption{Top: sulcal curves (red) are displayed on top of cortical surfaces. Bottom: sulcal curves are projected and smoothed in {the rectangular domain $[-\pi,\pi] \times [0,\pi]  \subset \mathbb{R}^2$.}}
	\label{fig:R2_sulcal_2subj}
\end{figure}

The concave regions in the highly convoluted cerebral cortex of the human brain are referred to as the {\em sulci} (Fig. \ref{fig:R2_sulcal_2subj}). These complex tree-shaped sulcal curves are highly variable in length, area, depth, curvature and topology across different subjects \cite{cachia.TMI.2003}. There have been extensive studies that connect the variabilities of such biomarkers with the differences in cognitive or pathological characteristics between populations \cite{im2019sulcal}. However, since each subject have different topological patterns, it is difficult to match the sulcal patterns across subjects \cite{huang.2020.TMI}. One approach of  reducing the difficulty of matching  is to smooth the sulcal patterns and then match the smoothed patterns. Such smoothed representations enable pattern matching in a continuous probabilistic fashion. Thus, we propose to use the Wasserstein distance, which minimizes the optimal transport cost of moving between probability distributions. 

The Wasserstein distances have been previously  applied in various imaging applications. \cite{Wang.2016.CVPR} computed the Wasserstein distance  using the hyperbolic metric for cortical brain morphometry.  \cite{Kolouri.2018} proposed the sliced Wasserstein distance in speeding up the computation in pattern recognition.  \cite{Oh.2020} derived a cycle-consistent generative adversarial network based on the optimal transport for  MRI reconstruction.  Despite the {advance} of various algorithms for Wasserstein distance based pattern matching \cite{Chartrand.2009,Kolouri.2018}, the rational of involving {smoothing} in such algorithms is unclear. In this paper, we work out the mathematical and implementation details on the Wasserstein distance on heat kernel smoothing, the kernel version of diffusion. In the experiment, we demonstrated the Wasserstein distance based registration reduces the variability of sulcal patterns across subjects.

\section{Methods}

\subsection{Sulcal pattern data}

We used the processed T1-weighted MRI of 456 subjects (age-matched 274 females and 182 males) in the Human Connectome Project (HCP) \cite{vanessen.2012}. The MRI were obtained using a Siemens 3T Connectome Skyra  scanner with a 32-channel head coil \cite{glasser2013minimal,smith.2013}. The MRI were registered to the MNI space with a FLIRT affine and FNIRT nonlinear registration \cite{jenkinson.2002}.  FreeSurfer's recon-all pipeline was used on the distortion- and bias-corrected MRI \cite{fischl2007cortical} that includes the segmentation of white matter and pial surfaces as well as  FreeSurfer's {folding-based} surface registration to the template. TRACE algorithm was used for automatic sulcal curve extraction from surface meshes \cite{huang.2020.TMI,lyu.2018} (Fig. \ref{fig:R2_sulcal_2subj}) and then projected on the unit sphere, which is further parameterized by spherical angles $p = (x, y) \in [-\pi,\pi] \times [0,\pi] \subset \mathbb{R}^2$. Since the data is spherical, it is periodic over $x$. Fig. \ref{fig:R2_sulcal_2subj} displays the sulcal curves of two different subjects on $\mathbb{R}^2$, where sulcal curves are assigned value  $1$ (colored red) {while all other parts of the brain} are assigned value 0.

\subsection{Heat kernel smoothing of sulcal patterns}
The sulcal pattern $f(p)$ is smoothed with heat kernel to reduce high frequency noise. {The} smoothing increases the signal-to-noise ratio (SNR) and increases the statistical power in the population study \cite{huang.2020.TMI}. We model the sulcal pattern as
\[
f(p)=\mu(p)+\varepsilon(p),\quad p=(x,y)\in [-\pi,\pi] \times [0,\pi] \subset \mathbb{R}^2,
\]
where $\mu(p)$ is the underlying true signal and $\varepsilon(p)$ is the noise. 
For observed data, we assign value 1 to the sulcal {curves} and 0 otherwise (Fig. \ref{fig:R2_sulcal_2subj}). We estimate $\mu(p)$ by smoothing with isotropic heat kernel. The smoothed estimate $\widehat{\mu}$ is given by the solution of the isotropic diffusion \[
	\frac{\partial}{\partial \sigma} u(x, y, \sigma) = \frac{\partial^2}{\partial^2 x} u(x, y, \sigma) + \frac{\partial^2}{\partial^2 y} u(x, y, \sigma)
\]
with the initial condition $u(x, y, \sigma=0)=f(x, y)$ and the boundary conditions
\[
\begin{aligned}
		&u\left(-\pi, y, \sigma\right)=u\left(\pi, y, \sigma\right), u^{\prime}\left(-\pi, y, \sigma\right)=u^{\prime}\left(\pi, y, \sigma\right),\\
	&u(x, 0, \sigma)=u\left(x, \pi, \sigma\right)=0.
	\end{aligned}
\]
The amount of smoothing $\sigma$ is determined empirically. The solution is given as the weighted Fourier series \cite{Chung.2007.WFS}: 
\begin{equation}
		\label{eq:R2hks}
		\widehat{\mu}(x,y)=	\sum_{j=0}^m \sum_{k=0}^{n}e^{-\lambda_{jk}\sigma}\bigg[A_{jk}\phi_{jk}^1+B_{jk}\phi_{jk}^2\bigg].
\end{equation}
The eigenvalues $\lambda_{jk}=j^2+ k^2$ and the eigenfunctions corresponding to Laplacian $\frac{\partial^2}{\partial^2 x} + \frac{\partial^2}{\partial^2 y}$ are 
\[
	\begin{aligned}
		\phi_{jk}^1 = \frac{2 \cos(jx)\sin(ky)}{\pi^2(1+ {\delta_{j0}})}, \phi_{jk}^2 = \frac{2 \sin(jx)\sin(ky)}{\pi^2},
	\end{aligned}
\]
where {$\delta_{j0}=1$ if $j=0$ and 0 otherwise.} The coefficients $A_{jk}$ and $B_{jk}$ are estimated  in the least square fashion using the initial data $f(p)$  \cite{Chung.2007.WFS}. The smoothing results are shown in Fig. \ref{fig:R2_sulcal_2subj} with diffusion time $\sigma = 0.001$.

{The} smooth representation enables alignment of different sulcal patterns on a common grid. For each subject, we first obtain the Fourier coefficients $A_{jk}$ and $B_{jk}$ via smoothing. The value on the grid point $(x_i,y_j)$ is then given by $\widehat{\mu}(x_i,y_j)$ {in} the expansion (\ref{eq:R2hks}). 

\subsection{Wasserstein distance on heat kernel smoothing}
Let $f_1$ {and} $f_2$ be the empirical distribution on the two sets of scatter points that define the vertices of sulcal curves 
$\mathbf{P}_{1}= \{p_1,\cdots,p_{n}\}$ and $\mathbf{P}_{2}=\{q_1, \cdots, q_{n}\}$:
\begin{equation}
	f_{1}(p)=\frac{1}{n} \sum_{i=1}^{n} \delta\left(p -p _{i}\right), \; f_{2}(q)=\frac{1}{n} \sum_{i=1}^{n} \delta\left(q-q_{i}\right).
	\label{eq:empirical_density}
\end{equation}
After  algebraic derivations  involving Choquest's and Birkhoff's theorems \cite{Peyre.2019}, the 2-Wasserstein distance $D_W$ between the empirical distributions is given as the Monge formulation
\begin{equation}
\begin{aligned}
D_W(f_1,f_2) =\inf_{\tau \in S_n}\Big(  \frac{1}{n}\sum_{i=1}^n \| p_i - p_{\tau(i)} \|^2   \Big)^{1/2},
\end{aligned}
\label{eq:DW_2}
\end{equation}
where $S_n$ is the permutation group of order $n$.

The diffused sulcal pattern can be written as the kernel convolution on initial condition $f(p)$ as $ \widehat{\mu}(p) = K_{\sigma} * f(p)$. 
with heat kernel $K_{\sigma}$ \cite{Chung.2007.WFS}. In $\mathbb{R}^2$, the heat kernel is simply Gaussian kernel 
\[
K_{\sigma}(p, p_i) = \frac{1}{2 \pi \sigma^2} e^{-\frac{1}{2\sigma^2}(p -p_i)^{\top}(p -p_i)}.
\]
The heat kernel smoothing on sulcal patterns ${\bf P}_1$ and ${\bf P}_2$ is given by (Fig. \ref{fig:R2_sulcal_2subj} Bottom)
\begin{equation}
	\label{eq:Gaussian_smoothed_density}
	\widetilde{f}_1(p) =\frac{1}{n} \sum_{i=1}^{n}K_{\sigma} (p, p_i), \; 
	\widetilde{f}_2(q) =\frac{1}{n} \sum_{i=1}^{n}K_{\sigma} (q, q_i).
\end{equation}
Let $\mathcal{A}(\widetilde{f}_1,\widetilde{f}_2)$ be the set of all possible joint density functions corresponding to $\widetilde{f}_1$ and $\widetilde{f}_2$. The Wasserstein distance between diffused sulcal pattern can be written as  \cite{Givens.1984}
\begin{equation}
	D_W(\widetilde{f}_1,\widetilde{f}_2)=\inf_{ \pi \in  \mathcal{A}(\widetilde{f}_1,\widetilde{f}_2)} \Big(  \int \|p -q \|^2 \pi(p,q)  \mathrm{d}p \mathrm{d}q \Big)^{\frac{1}{2}}.
	\label{eq:DW_Gaussian_inter1}
\end{equation}
If we restrict the joint distribution to  be {a} linear combination of multivariate normals, we can compute the expression (\ref{eq:DW_Gaussian_inter1}) exactly. We will denote the Wasserstein distance with such a restriction as $D_{W'}$. Then we have the following equivalence.  

	\begin{figure}[t]
	\centering
	\includegraphics[width=1\linewidth,clip=true]{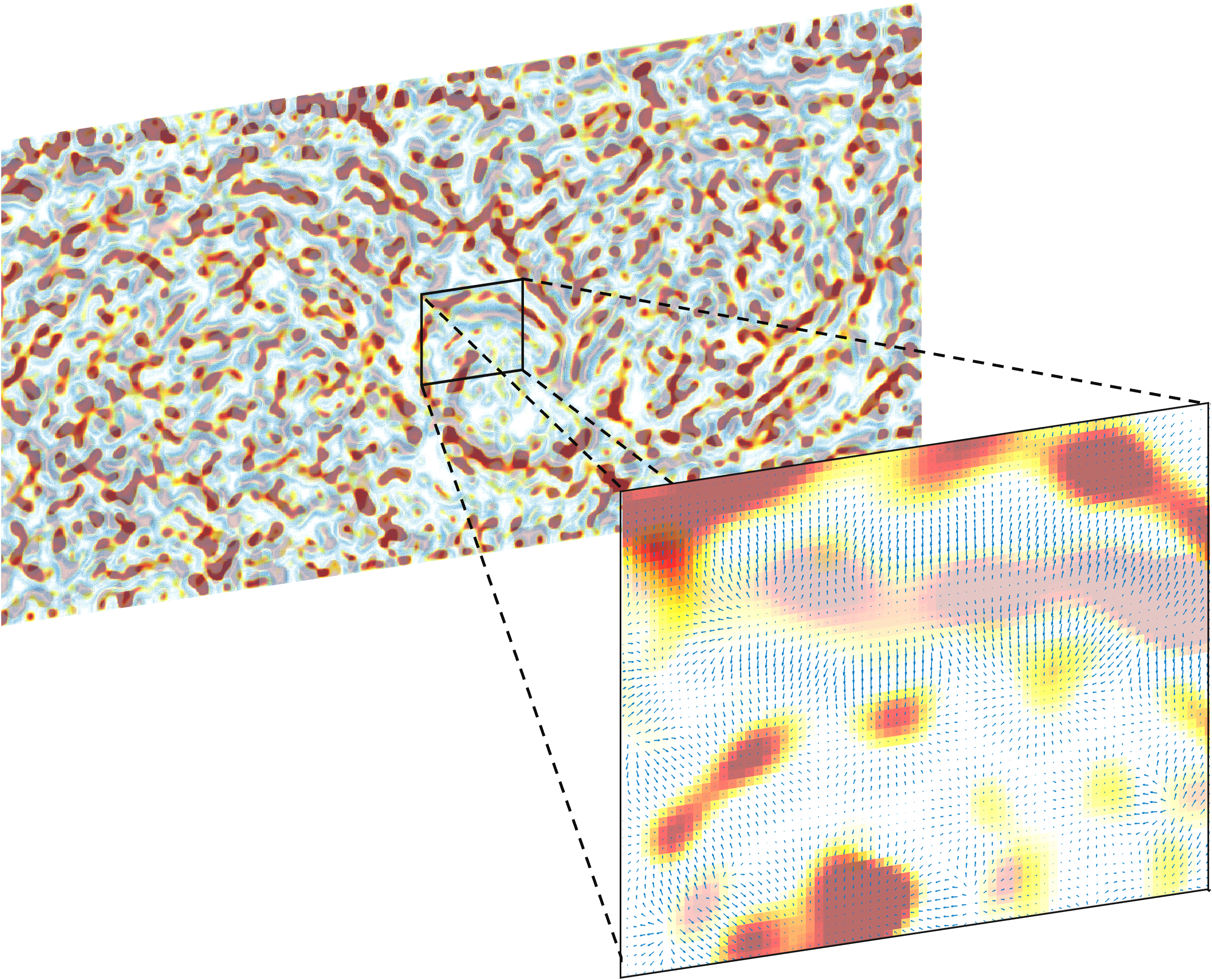}
	\caption{Subject 1 (lower transparency) superimposed on top of subject 2 (higher transparency). The arrows are the displacement field of matching subject 1 to subject 2.}
	\label{fig:R2_sulcalONLY_deform}
\end{figure}

\begin{theorem} 
	For heat kernel smoothing $\widetilde{f}_1$ and $\widetilde{f}_2$, we have 
		\begin{equation}
		D_{W'}(\widetilde{f}_1,\widetilde{f}_2)=D_W(f_1,f_2). \label{eq:invariance}
	\end{equation}
\end{theorem}
{\em Proof.} We provide the sketch of proof. The joint density $\pi$ in \eqref{eq:DW_Gaussian_inter1} can be expressed as the mixture of multivariate normals as
$\pi(p, q)=\sum_{i, j=1}^n \pi_{i j} \phi_{i j}(p, q),$
{where the mixing proportions $\pi_{ij}$ are doubly stochastic and satisfies $\sum_{i}\pi_{ij}=\sum_{j}\pi_{ij}=1$. The multivariate normals $\phi_{i j}$ follow
\[
\phi_{i j} \sim 
	{N} \left(\begin{bmatrix}
		p_i \\
		q_j
		\end{bmatrix},\begin{bmatrix}
		\sigma^2 I_2 & H \\
		H^{\top} & \sigma^2 I_2
		\end{bmatrix}\right),	
\]}
where $p_i\in \mathbf{P}_1$, $q_i\in \mathbf{P}_2$, $I_2$ is the identity matrix and $H$ is the $2\times 2$ {positive semi-definite covariance matrix. Then after lengthy derivations,} we can obtain
\[
	D_{W^{\prime}}(\widetilde{f}_1, \widetilde{f}_2)^2=\inf _{\pi_{i j}, H}  \sum_{i, j=1}^n \pi_{i j}\left\| p_i- q_j\right\|^2+2 \operatorname{tr} (\sigma^2 I_2-H ).
\]
Subsequently, it suffices to minimize each term in $D_{W^{\prime}}$ separately. Minimizing the first term leads to the same result as \eqref{eq:DW_2} while the second term results in zero \cite{Givens.1984}. \hfill{$\square$}

Since $D_{W}(\widetilde{f}_1,\widetilde{f}_2) \leq D_{W'}(\widetilde{f}_1,\widetilde{f}_2)$, we have
$$D_{W}(\widetilde{f}_1,\widetilde{f}_2) \leq D_W(f_1,f_2)$$ indicating that heat kernel smoothing 
{reduces} the Wasserstein distance between the sulcal patterns and their variabilities.

\subsection{Gradient descent on the dual formulation}

Given two {smooth patterns} $\widetilde{f}_1$ and $\widetilde{f}_2$, the computation of the Wasserstein distance involves mapping a point $p$ from the first pattern to the corresponding point $U(p)$ in the second pattern (Fig. \ref{fig:R2_sulcalONLY_deform}). {The deformation $p \to U(p)$ is required to have the smallest possible total displacement and satisfies}
\[
	\int_{U^{-1}(A)} \widetilde{f}_1(p)\mathrm{d} p=\int_A \widetilde{f}_2(q)\mathrm{d} q, \quad \forall A\subset\mathbb{R}^2.
\]
This leads to solving the optimization problem
\begin{equation}
	\label{eq:DWmatching_orig}
	\inf_U \int  \|U(p) -p \|^2 \widetilde{f}_1(p)\mathrm{d}p, 
\end{equation}
which can be further reduced to minimizing 
\begin{equation}
	\label{eq:DWmatching_1}
	L(\varphi,\psi) = \int \varphi(p) \widetilde{f}_1 (p)\mathrm{d}p+\int \psi(q) \widetilde{f}_1 (q)\mathrm{d}q
\end{equation}
with constraint $\varphi(p)+\psi(q)\geq  {p^{\top}q}$  \cite{Chartrand.2009}. 
\begin{theorem} \cite{Knott.1984}
	The unique solution $(\varphi,\psi)$ to the problem \eqref{eq:DWmatching_1} with constraint $\varphi(p)+\psi(q)\geq p^{\top}q$ are convex conjugates, i.e., $\psi=\varphi^*$ and $\varphi=\psi^*$ where $\varphi^*$ and $\psi^*$ are defined as
	\[
		\begin{cases}
			\varphi^*(q)=\max _{p }(p^{\top}q-\varphi(p))\\
			\psi^*(p)=\max_{q}(q^{\top}p-\psi(q))
		\end{cases}.
	\]
The solution to \eqref{eq:DWmatching_orig} is then given by $U=\nabla \varphi.$
\end{theorem}
{The  solution is given implicitly. However, the derivative of \eqref{eq:DWmatching_1} can be explicitly found.}
\begin{theorem} \cite{Chartrand.2009}
{Suppose $\varphi$ is the convex conjugate of $\psi$ for some $\psi$ and $\nabla^2 \varphi$ is Lipschitz.  Then it follows $L$ is convex, Lipschitz and
	\[
		L'(\varphi)= \widetilde{f}_1 -  (\widetilde{f}_2\circ \nabla \varphi)\det(H_{\varphi}),
	\]
	where $H_{\varphi}$ is he Hessian matrix of $\varphi$.}
	\label{thm:gd}
\end{theorem}
Subsequently,  we solve for gradient descent on $\varphi$ as \cite{Chartrand.2009}:
\begin{equation}
	\label{eq:gd}
	\varphi_{n+1} = \varphi_n -  L'(\varphi_n),
\end{equation}
The overall computational complexity for the algorithm is $O(N)$ \cite{Kolouri.2017}, which is more scalable than the Hungarian algorithm with $O(N^3)$ \cite{Munkres.1957} and fluid mechanics with $O(N^{3/2})$  \cite{Benamou.2000}. In this study, $800\times 400$ rectangular grid {with $N=320000$ vertices were used. The grid values {$\widehat \mu(x_i, y_j)$  serve} as the input for the gradient descent.}  We display the result of the gradient descent on estimating the displacement field $\nabla\varphi (p)- p$ as arrows in Fig. \ref{fig:R2_sulcalONLY_deform}. Due to the convexity of $\varphi$, the algorithm is expected to recover the global optimum (Remark 3.3 in \cite{Chartrand.2009}). 

\begin{figure}[t]
	\centering
	\includegraphics[width=1\linewidth,clip=true]{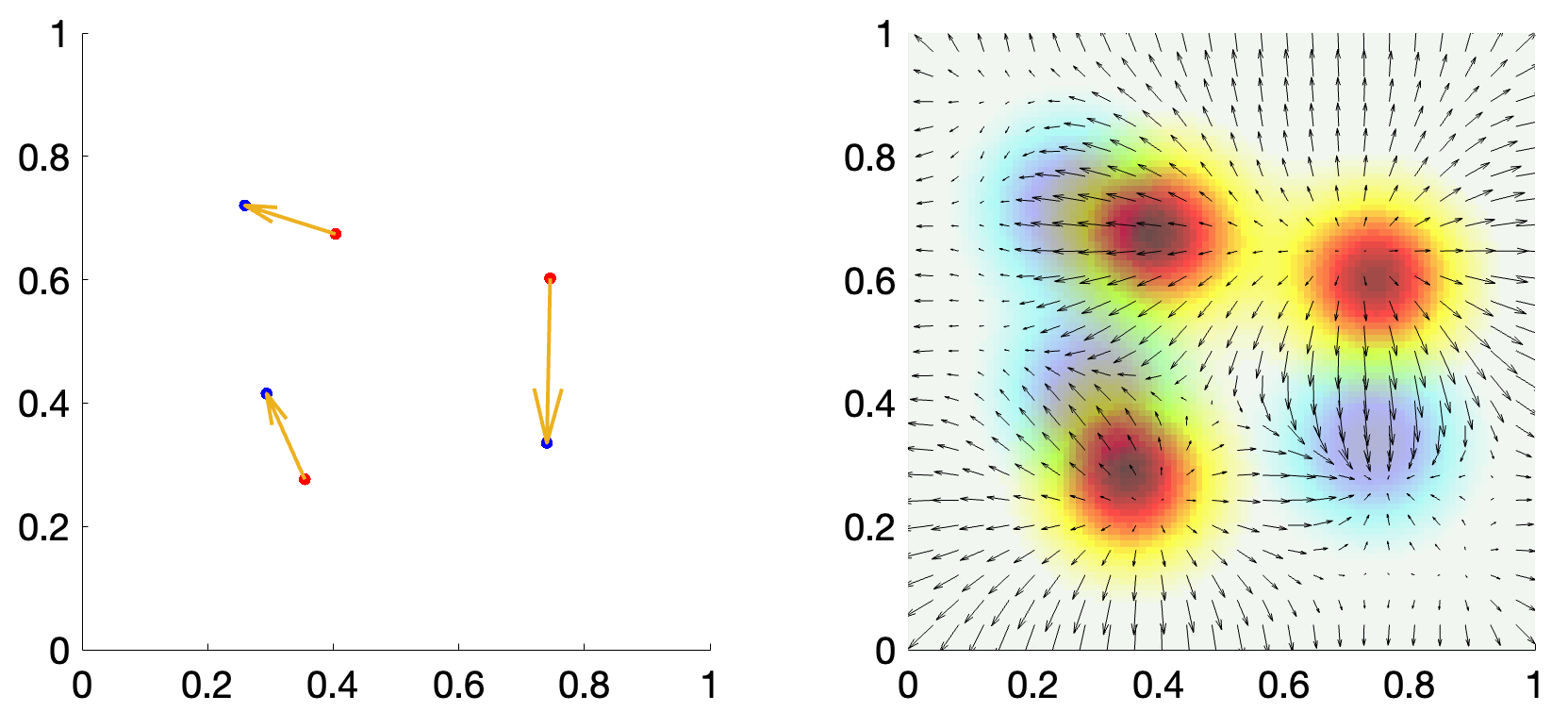}
	\caption{Left: the Hungarian Algorithm is used to match two sets of three  points (colored red and blue). Right: the gradient descent is used to match the heat kernel smoothing on these points. Arrows are the estimated displacement $U(p)-p$.}
	\label{fig:validation}
\end{figure}

\section{Experiments}

\subsection{Validation against the Hungarian Algorithm}

We matched the two sets of $n$ random generated scatter points in $[0,1]^2$. Fig.\ref{fig:validation} displays the result of one realization with $n=3$. The Wasserstein distance $D_W(f_1, f_2)$  can be computed exactly through \eqref{eq:DW_2} using the Hungarian Algorithm \cite{Munkres.1957}. We then applied heat kernel kernel smoothing with bandwidth $\sigma = 0.01$ on the scatter points and compute the Wasserstein distance $D_W'(\tilde f_1, \tilde f_2)$ using the gradient descent \eqref{eq:gd}. The average percentage reduction of distance for 100 independent simulations for $n=2,3,4$ are $0.86 \pm 0.05$, $0.84 \pm 0.06$ and $0.77 \pm 0.04$ respectively. This is consistent with our theoretical result in section 2.3. The computer code for performing Wasserstein distance based matching is provided in \url{https://github.com/laplcebeltrami/sulcaltree}.

\begin{figure}[t]
	\centering
	\includegraphics[width=1\linewidth,clip=true]{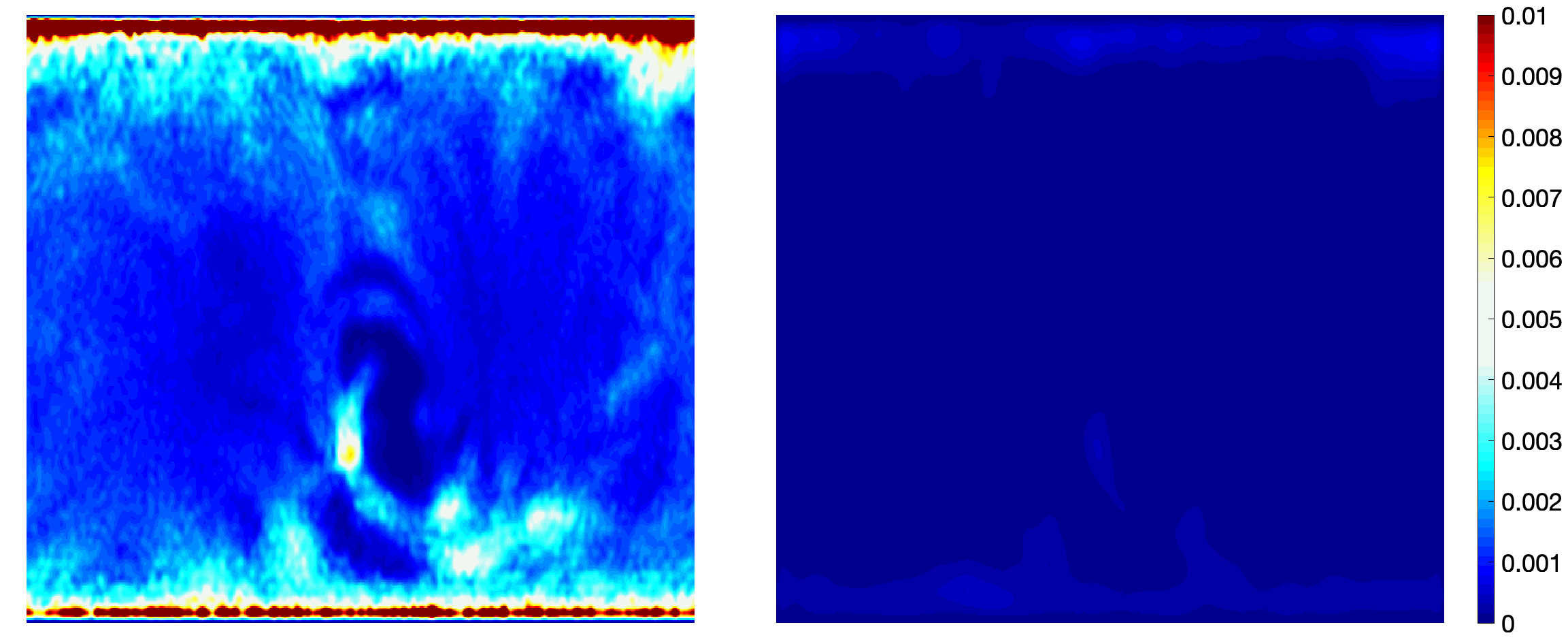}
	\caption{{The intersubject variability  of the {smoothed} original sulcal patterns (left) and deformed patterns (right).}}
	\label{fig:variablity}
\end{figure}

\subsection{Reduction of image registration variability}

The sulcal pattern has been already nonlinearly aligned using FreeSurfer's standard folding-based surface registration to the surface atlas \cite{fischl2007cortical}. Wasserstein distance based matching significantly reduces the sulcal pattern variability further from the  FreeSurfer alignment. We first compute the average of smoothed sulcal pattern maps  $T = \frac{1}{n}\sum_{i=1}^n \widetilde{f}_i  (p)$ on smoothed estimates \eqref{eq:R2hks}. This serves as the reference template pattern for subsequent registration. We then align each subject to the average pattern $T$ using  gradient descent \eqref{eq:gd}. We initialize $\varphi$ with $\frac{1}{2}\|p\|^2$ so that initial mapping $\nabla \varphi$ becomes the identity. Such initialization is data-free and has demonstrated efficacy in various applications including facial \cite{Kolouri.2017} and brain images \cite{Chartrand.2009}. {We set as the maximum number of iterations at 200.} The average runtime for one subject is approximately 8 seconds on a desktop computer.

For deformation $U_i(x)$ for subject $i$, the deformed sulcal patterns are obtained as $\widetilde{f}_i^T (p) = \widetilde{f}_i (U_{i}(p))$. The statistical variability before registration is computed as the sample variance of sulcal patterns  $\widetilde{f}_i  (p)$ at each point across subjects (Fig. \ref{fig:variablity}). The variability is significantly reduced after deformation. The mean variability over all the points in the brain is 0.0021 for the {smoothed} original pattern while it is $7.8\times 10^{-5}$, the significant reduction of variability by $96.29\%$. 

\subsection{Sexual dimorphism}
\begin{figure}[t]
	\centering
	\includegraphics[width=1\linewidth,clip=true]{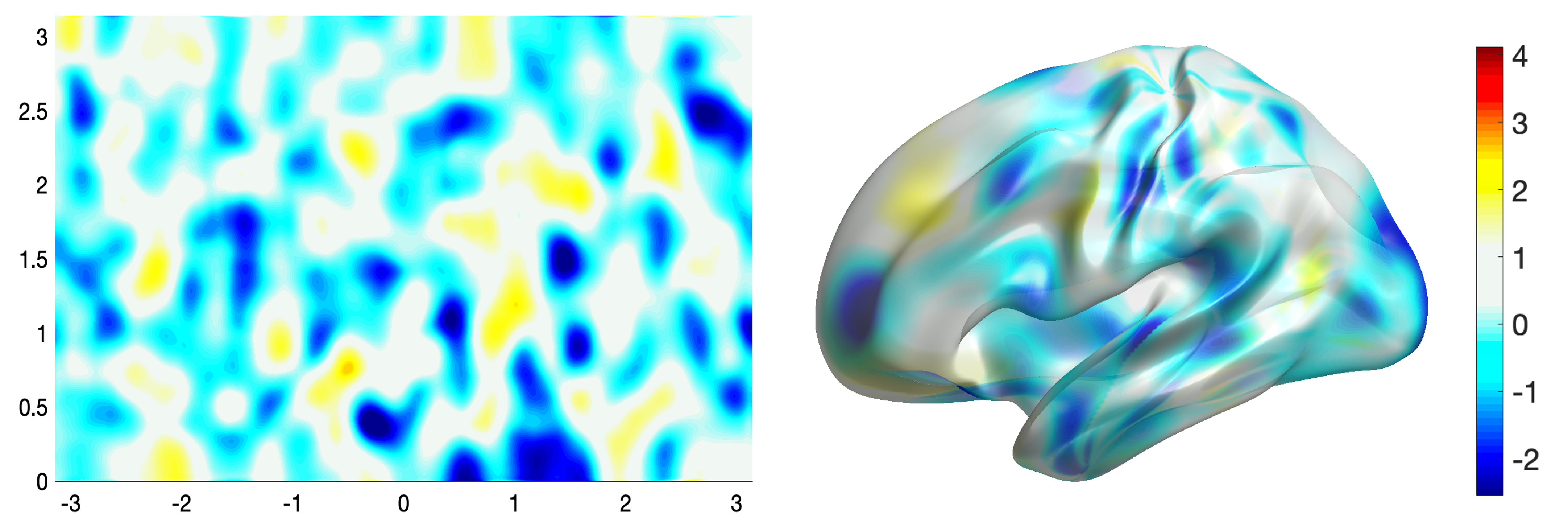}
	\caption{$t$-statistics map of the localized sulcal pattern differences (female-male). Values are thresholded in $[-2.84,4.12]$ corresponding to the corrected $p$-value of 0.05. {Most of female brain regions show less sulci, which corresponds to more gray matter and thicker cortical thickness observed in other studies.}}
	\label{fig:sec4_t}
\end{figure}

The method is subsequently used to determine the sulcal pattern differences between  240 females and 172 males. The two sample $t$-statistics were constructed on the deformed smoothed sulcal patterns $\widetilde{f}_i^T (p) = \widetilde{f}_i (U_{i}(p))$   (Fig. \ref{fig:sec4_t}). The threshold $t$-statistic value greater than $4.12$ or smaller than $-2.84$ correspond to the $p$-value of 0.05 after multiple comparisons correction through the permutation test (half millions)  for the deformed pattern \cite{chung.2019.CNI}. 
We obtained numerous significant differences all over the brain regions including the temporal lobe, which is responsible for sensory processing and known for sex difference \cite{huang.2020.TMI}. {The negative $t$-statistics in most of the brain regions indicates the presence of less sulci for females. This seems to be related to findings in \cite{luders.2009}, where females have larger gray matter volumes in a number of areas including the left temporal gyrus. In \cite{im.2006}, cortical thickening is found extensively in the left superior parietal gyrus and postcentral gyrus, which is consistent with less sulci in our study.}

\section{Discussion}

In this paper, we presented the new framework of matching sulcal patterns of human brain across subjects. We provided the theoretical justification for performing heat kernel smoothing before computing the Wasserstein distance. Smoothing reduces the Wasserstein distance between the sulcal patterns and spatial pattern variabilities. It is also possible to further refine the registration performance via the multi-resolution on heat kernel smoothing \cite{Chartrand.2009} or {iteratively using} 
the average of the deformed sulcal patterns $\frac{1}{n}\sum_{i=1}^n\widetilde{f}_i^T (p)$ as the template in the next iteration. These are left as future studies.

The FreeSurfer uses the irregular triangle meshes while our smoothing and the gradient descent uses the regular pixel grid. This may introduce potential interpolation artifacts. A better approach would be to perform smoothing and the gradient descent on triangle meshes. This is left as a future study.
  
\section{Acknowledgement}
We would like to thank Soheil Kolouri of Vanderbilt University and Tahmineh Azizi of University of Wisconsin-Madison for discussion on the Wasserstein distance, and Ilwoo Lyu of Ulsan National Institute of Science and Technology for assistant with the TRACE algorithm.

\bibliographystyle{IEEEbib}
\bibliography{reference.ISBI.2023_new}
\end{document}